\documentclass[aoas,nameyear,MSNbibl,dvips]{arximspdf}
\usepackage{stfloats}
\usepackage{graphicx}

\doi{10.1214/10-AOAS398REJ}
\referstodoi{10.1214/10-AOAS398}
\volume{5}
\issue{1}
\pubyear{2011}
\firstpage{99}
\lastpage{123}

\makeatletter
\renewcommand{\epsilon}{\varepsilon}
\newcommand{\iid}{\stackrel{\mathrm{i.i.d.}}{\sim}}
\fnbelowfloat
\makeatother

\begin{document}
\begin{frontmatter}

\title{Rejoinder}
\runtitle{Rejoinder}
\pdftitle{Rejoinder on A statistical analysis of multiple temperature proxies:
Are reconstructions of surface temperatures over the last 1000 years reliable?
by B. B. McShane and A. J. Wyner}

\begin{aug}
\author[A]{\fnms{Blakeley B.} \snm{McShane}\ead[label=e1]{b-mcshane@kellogg.northwestern.edu}%
\ead[label=u1,url]{http://www.blakemcshane.com}}
\and
\author[B]{\fnms{Abraham J.} \snm{Wyner}\corref{}\ead[label=e2]{ajw@wharton.upenn.edu}%
\ead[label=u2,url]{http://statistics.wharton.upenn.edu}}
\runauthor{B. B. McShane and A. J. Wyner}
\affiliation{Northwestern University and the University of Pennsylvania}
\address[A]{Northwestern University \\
Kellogg School of Management \\
Leverone Hall \\
2001 Sheridan Road \\
Evanston, Illinois 60208 \\
USA\\
\printead{e1}\\
\printead{u1}} %adresu isvedimo komanda gale!
\address[B]{University of Pennsylvania \\
The Wharton School \\
400 Jon M. Huntsman Hall \\
3730 Walnut Street \\
Philadelphia, Pennsylvania 19104 \\
USA\\
\printead{e2}\\
\printead{u2}}
\end{aug}

% HISTORY:
\received{\smonth{11} \syear{2010}}
\revised{\smonth{12} \syear{2010}}

% ABSTRACT

% KEYWORDS
%
%

\end{frontmatter}

We heartily thank Michael Stein and Brad Efron for selecting our paper
for discussion and for the tremendous task of recruiting and editing 13
discussion articles on this controversial and timely topic. We are
grateful for the opportunity to receive feedback on our work from such
a large number of knowledgable discussants whose work and fields of
expertise are so broad. The fact that our paper was of interest not
only to academic statisticians and climate scientists but also
economists and popular bloggers\footnote{Steve McIntyre of
\href{http://www.climateaudit.org}
{www.climateaudit.org} and Gavin Schmidt of \href{http://www.realclimate.org}
{www.realclimate.org}.} bespeaks the importance of the topic.

We thank all 13 discussants for the time they put into considering and
responding to our paper. Each one deserves a detailed response, but
time and space constraints make that impossible. We therefore
acknowledge the favorable tenor of the discussions generally if not
specifically.

The discussion has great value, particularly for raising points of
contrast sometimes about fundamental issues. For instance, Wahl and
Amman (WA) note ``there is an extensive literature contradicting
\citeauthor{McSWyn11}'s (\citeyear{McSWyn11}) assertions about low or poor
relationships between proxies
and climate''. On the other hand, Tingley asserts ``each proxy is weakly
correlated to the northern hemisphere mean (for two reasons: proxies
generally have a~weak correlation with local climate, which in turn is
weakly correlated with a hemispheric average)'' and Davis and Liu (DL)
state ``there is just not much signal present''. This contrast can be
explained at least in part by context. Our paper addresses the specific
task of reconstructing annual temperatures over relatively short epochs
during which temperatures vary comparatively little. Nevertheless, such
contrasts are suggestive of the important frontiers for research and we
hope our paper and this discussion will lead to advances on these fronts.

In this rejoinder, we aim to do three things. First, we respond to the
detailed and highly critical discussion of Schmidt, Mann, and
Rutherford (SMR). Next, we reiterate our key findings while targeting
themes that emerge from multiple discussants. Finally, we conclude with
a more in-depth response to Tingley and Smerdon who address the same
broad issue. The discussions of SMR and Tingley are noteworthy because
they take a ``scientific'' approach as opposed to the ``statistical''
approach taken by many of the other discussants (e.g., DL and Kaplan),
thereby highlighting some of the differences between various approaches
to data analysis [\citet{Diac85}] and pointing to some of the weaknesses
of the former in high uncertainty settings such as proxy-based global
temperature reconstruction. Smerdon's discussion, on the other hand,
heeds both scientific and statistical considerations.\looseness=-1

Some of the discussants chose to highlight questions and problems
related to the introduction and history [Nychka and Li (NL), WA].
Others reflected on approaches outside the scope of our expertise (H\"
{o}lmstrom). While these are interesting topics, they are also
tangential to the central issue of our paper---the \textit{uncertainty} of
proxy-based global temperature reconstructions (i.e., the second moment
rather than the first). In our rejoinder we will focus more narrowly on
this topic.

This short form version of the rejoinder should be read as a summary
document. A fuller version, which contains the supporting details and
figures for the claims made here, can be found as a supplementary
information (SI) document at the \textit{Annals of Applied Statistics}
supplementary materials website [\citet{McSWyn11r1}]. The short and long
documents are divided into the same sections for easy reference and the
reader interested in the full treatment can read the long document
alone without reference to the short one. As with our paper, code to
implement all analyses conducted for the rejoinder is available at the
\textit{Annals of Applied Statistics} supplementary materials website
[\citet{McSWyn11r2}].

%s1 ###
\section{Rejoinder to SMR}\label{smr}

Broadly, SMR engage in a two-fold critique of our conclusions. First
(SMR Figure 1), they aim to show that our 1000-year temperature
reconstructions based on real proxy data are flawed, allegedly because
we miss important problems in a subset of the data. Second (SMR Figure
2), they argue through a simulation study that the RegEM EIV
(Regularized Expectation-Maximization Errors-In-Variables Algorithm,
referred to throughout this rejoinder as RegEM, RegEM EIV, and EIV)
method is vastly superior to the methods examined and applied in our
paper. We show that this is not true.

Before embarking on our discussion of their work, we must mention that,
of the five discussants who performed analyses (DL, Kaplan, SMR,
Smerdon, and Tingley), SMR was the only one who provided an incomplete
and generally unusable repository of data and code. The repository
created by SMR \textit{specifically for this discussion} was, like that of
the other four discussants, graciously provided and quite usable.
However, we lacked clear and easily implementable code (i) to fit RegEM
EIV ourselves and (ii) to draw new temperatures and pseudoproxies from
their simulation model. Code for these purposes is archived by Mann at
\href{http://www.meteo.psu.edu/\textasciitilde  mann/PseudoproxyJGR06/}
{http://www.meteo.psu.edu/\textasciitilde  mann/}
\href{http://www.meteo.psu.edu/\textasciitilde  mann/PseudoproxyJGR06/}{PseudoproxyJGR06/}
and
\href{http://www.meteo.psu.edu/\textasciitilde  mann/supplements/MultiproxyMeans07/}
{http://www.meteo.psu.edu/\textasciitilde  mann/supplements/}
\href{http://www.meteo.psu.edu/\textasciitilde mann/supplements/MultiproxyMeans07/}
{MultiproxyMeans07/}.

Among other things, the RegEM EIV fitting procedure cannot be executed
by a straightfoward function call as is typical for statistical code
libraries. Rather,\ the archives consist of a large number of files
layered on top of one  another and, despite a major effort on our part,
we were unable to replicate  published results within the publication
time constraints of this rejoinder. That said, independent researchers
have, after important modifications,\footnote{A bypass of the function
used to generate new pseudoproxies during each run (pseudoproxytwo.m)
is required since this module appears to be inoperative.} successfully
run code from the first URL (Jason Smerdon, personal communication).
Consequently, throughout this section, we work with RegEM $\hat y$'s
(pre-fit by SMR to both real and simulated data) as well as one
particular draw of the data from their simulation which were
provided.

%s1.1 ###
\subsection{Proxy data: Full versus reduced network of 1000 year-old
proxies}\label{smr:proxy}

SMR allege that we have applied the various methods in Sections~\ref{sec4}
and~\ref{sec5}
of our paper to an inappropriately large group of 95 proxies which date
back to 1000 AD (93 when the Tiljander lightsum and thicknessmm series
are removed due to high correlation as in our paper; see footnote 11).
In contrast, the reconstruction of \citet{Mannetal08} is applied to a
smaller set of 59 proxies (57 if the two Tiljander series mentioned
previously are removed; 55 if all four Tiljander series are excluded
because they are ``potentially contaminated'').

The process by which the complete set of 95$/$93 proxies is reduced to
59$/$57$/$55 is only suggestively described in an online supplement to \citet
{Mannetal08}.\footnote{The \citet{Mannetal08} Supplementary Information
contains the following note: ``Tree-ring data included 926 tree-ring
series extracted from the International Tree Ring Data Bank (ITRDB,
version 5.03: \href{http://www.ncdc.noaa.gov/paleo/treering.html}{www.ncdc.noaa.gov/paleo/treering.html}).
All ITRDB tree-ring proxy series were required to pass a series of minimum
standards to be included in the network: (i) series must cover at least
the interval 1750 to 1970, (ii) correlation between individual cores
for a given site must be 0.50 for this period, (iii) there must be at
least eight samples during the screened period 1800--1960 and for every
year used''.} As statisticians we can only be skeptical of such
improvisation, especially since the instrumental calibration period
contains very few independent degrees of freedom. Consequently, the
application of  ad hoc  methods to screen and exclude data
increases model uncertainty in ways that are unmeasurable and uncorrectable.

\begin{figure}[t]

\includegraphics{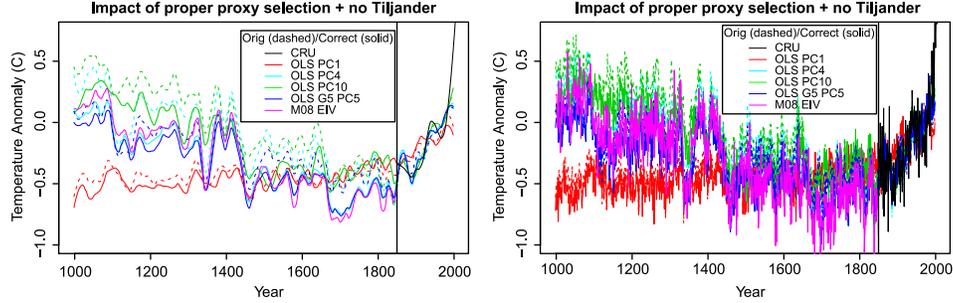}

\caption{Reproduction of SMR Figure 1a. The left panel gives smoothed
fits thereby reproducing SMR Figure 1a whereas the right panel gives
unsmoothed fits. Results using the reduced set of $55$ Mann et al.
(\protect\citeyear{Mannetal08}) proxies (excluding Tiljander) are plotted with solid
lines whereas results using the full set of $93$ proxies are plotted
with dashed lines. Two features stand out from these plots. First, the
differences between the fit of a given method to the full or reduced
set of proxies are quite small compared to the annual variation of a
given fit or compared to the variations between fits. Second, the RegEM
EIV methods produce reconstructions which are nearly identical to those
produced by OLS $\mathit{PC}4$ and OLS $G5$ $\mathit{PC}5$. Compare also with SMR Figure
$S2$ which is similar to the bottom panel but excludes RegEM EIV.}
\label{fig:smr1a}
\end{figure}

Moreover, our interpretation of SMR Figure 1 is quite different. We see
the variation between the larger and smaller datasets as relatively
small with respect to the variation among the models. The appearance of
a difference in SMR Figure~1a is especially magnified because those
reconstructions are smoothed. Smoothing exaggerates the difference and
requires careful adjustment of fit statistics such as standard errors,
adjustments which are lacking in SMR and which are in general known
only under certain restrictive conditions. In contrast, consider the
right panel of Figure \ref{fig:smr1a} which is a reproduction of SMR
Figure~1a without smoothing. The difference between a given model fit
to the full dataset or the reduced data set is clearly dwarfed by the
annual variation of the fit; the full and reduced set of proxies yield
inconsequentially different reconstructions. It thus seems to us the
main point of Figure 14 of the paper (which SMR Figures 1a and S2
roughly mimic) stands: various methods which have similar holdout RMSEs
in the instrumental period produce strikingly different reconstructions
including\vadjust{\goodbreak} ``hockey sticks'' (such as the red one in Figure \ref
{fig:smr1a}), ``inverted check marks'' (such as the green), and things
in between (such as the blue and purple). In short, while SMR allege
that we use the ``wrong'' data, the result remains the same (also see
SI).

%f1 ###

%f2 ###
\begin{figure}

\includegraphics{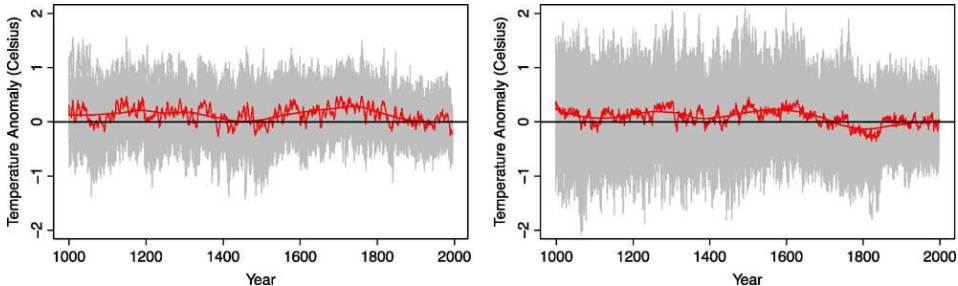}

\caption{Difference between our Bayesian $\mathit{AR}2+\mathit{PC}10$ model of Section
$5$ and various other models. The left panel gives the difference
between our Bayesian $\mathit{AR}2+\mathit{PC}10$ model fit to the network of $93$
proxies dating back to $1000$ AD and the original Mann et al.
(\protect\citeyear{Mannetal08})
RegEM EIV fit to the network of $59$ proxies dating back to $1000$ AD.
The right panel gives the difference between our Bayesian $\mathit{AR}2+\mathit{PC}10$
model fit to the network of $93$ proxies dating back to $1000$ AD and
the model of SMR Figure $1b$ (i.e., the same Bayesian $\mathit{AR}2+\mathit{PC}10$ model
but fit to the network of $55$ proxies dating back to $1000$ AD instead
of the network of $93$ proxies). As can be seen, there are no
statistically significant differences between these two models and our
Bayesian $\mathit{AR}2+\mathit{PC}10$ model fit to the $93$ proxies. The annual
difference is given in red, the smoothed difference in thick red, and
annual uncertainties bands are given in gray. The right plot has wider
intervals because the uncertainty in both models is accounted for.
Since we lack uncertainty estimates for RegEM, the left panel uses only
the uncertainty estimates of our Bayes model. Compare to SMR Figures
1b and 1c as well as figures in the SI.}
\label{fig:smr1b}
\end{figure}

We have two additional findings. First, as shown in Figure \ref
{fig:smr1a}, the RegEM reconstruction is \textit{nearly identical} to to
OLS PC4 and OLS G5 PC5. This is particularly interesting in light of
the performance comparisons of SMR Figure 2. Second, SMR Figure 1a and
our
Figure \ref{fig:smr1a} given here do not account for uncertainty in the model
fits. When such uncertainty is accounted for, as can easily be done for
the models in SMR Figures 1b and 1c, we see that the difference between
the reconstructions produced from the larger data set of 95$/$93 proxies
and the 59$/$57$/$55 are negligible with respect to overall uncertainty
(see Figure \ref{fig:smr1b}; see SI for more details).

%s1.2 ###
\subsection{The selection of principal components}\label{sec1.2}

SMR Figure 1c replots our Bayes model (Figure 16 of the paper) with two
differences: it uses the reduced dataset of 55 proxies and only four
principal components. There are no statistically significant
differences between the resulting model and our original one (see SI),
yet SMR allege that ``$K=10$ principal components is almost certainly
too large, and the resulting reconstruction likely suffers from
statistical over-fitting. Objective selection criteria applied to the
\citet{Mannetal08} AD 1000 proxy network, as well as independent
``pseudoproxy'' analyses discussed below, favor retaining only $K=4$''.

SMR are wrong on two counts. First, the two ``objective'' criteria they
suggest select differing numbers of principal components. Second, each
criterion has multiple implementations each producing different
results. As is well known to statisticians, there is no single
objective way to resolve these discrepancies. Furthermore, the PC
selection procedures that SMR prefer select ``significant'' PCs based
entirely on the matrix of predictors without considering the response
variable. To protect against overfitting, the selection process should
in some way take into account the relationship between the predictor
and the response [see also \citet{Izen08}, \citet{HasTibFri09}]. Compounding
matters, SMR implement their allegedly objective criteria in
nonstandard and arbitrary ways and several times in error.\footnote
{They appear to mistake the squared eigenvalues for the variances of
the principal components which leads to a thresholding of total
variance \textit{squared} instead of variance. We provide complete details
in the SI.} When correctly implemented, the number of principal
components retained varies across each ``objective'' criterion from two
to fifty-seven. Using ten principal components, therefore, can hardly
be said to induce the ``statistical over-fitting'' claimed by SMR.

%s1.3 ###
\subsection{Simulated data}\label{smr:sim}

SMR Figure 2 (along with SMR Table S1) purports to show that the Lasso
(applied in Section~\ref{sec3} of our paper) and the variety of principal
component methods (applied in Section~\ref{sec4}) are fundamentally inferior to
the RegEM EIV method of \citet{Mannetal08} and to thereby challenge our
assertion that various methods perform similarly (based on Figures
11--13 of the paper). RegEM is considered to be a state of the art
model for temperature reconstructions in the climate science literature
[\citeauthor{MaRuWaAm07}  (\citeyear{MaRuWaAm07,Mannetal08}), \citet{LeeZwiTas08}].

\begin{figure}

\includegraphics{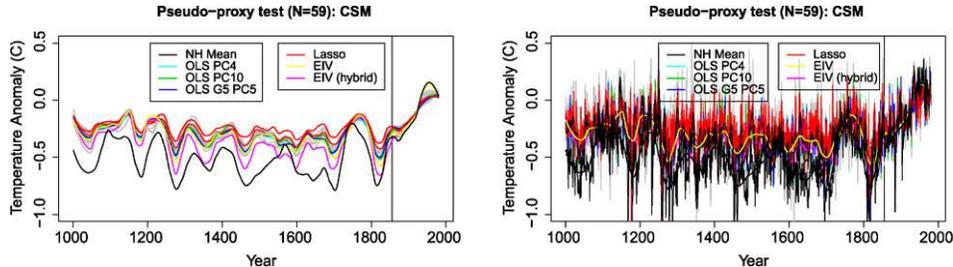}

\caption{Corrected version of SMR Figure~2a. The left panel gives
smoothed fits as in SMR Figure~2a whereas the right panel gives
unsmoothed fits. The annual variation of each method's fit dwarfs the
difference between methods. See SI for corrected versions of SMR
Figures 1b, 1c, and 1d.}\label{fig:smr2a}
\end{figure}

SMR Figure 2 is based on data simulated from National Center for
Atmospheric Research (NCAR) Climate System Model (CSM) as well as
Helmholtz-Zentrum Geesthacht Research Centre (GKSS) European Centre
Hamburg Ocean Primitive Equation-G (ECHO-G) ``Erik'' model. We see
several problems with this simulation:
\begin{longlist}[(3)]
\item[(1)] While we can vaguely outline the process which generated the
simulated temperatures and pseudoproxies, the details are buried in
layers of code at various supplementary websites and therefore are not
reproducible.
\item[(2)] In contrast to the methods of Sections~4 and~5 of our paper which
are transparent, RegEM appears to be a classic, improvised methodology
with no known statistical properties, particularly in finite samples or
when assumptions are violated. For instance, the ``missing at random''
assumption [\citet{LitRub02}] likely fails to hold here [\citet
{SmeKapCha08}]. Further, there are enormous numbers of variations on
RegEM (e.g., RegEM-Ridge, RegEM-Truncated Total Least Squares, etc.)
each with their own associated tuning parameters and no firmly agreed
upon methods for tuning them [\citet
{SmeKapCha08}, Christiansen, Schmith and Thejll (\citeyear{ChrSchThe09,ChrSchThe10}),
\citet{Ruthetal10}]. Consequently, we
cannot rule out the possibility that RegEM was tailor-made to the
specific features of this simulation,\footnote{This is suggested by the
fact that RegEM performs nearly identically to OLS PC4 and OLS G5 PC5
on the real proxy data (see SMR Figure 1a and our Figure \ref
{fig:smr1a}; see also SMR Figure 2 and our corrected versions in Figure
\ref{fig:smr2a} and the SI) but substantially better on the simulated
data (see SMR Table S1 and our corrected version in the SI).}
particularly since the same simulations have been used in repeated
studies.\footnote{For a review of papers using these simulations, see
\citet{SmeKapCha08} who state in their opening two paragraphs: ``\citet
{Ruthetal05} used RegEM to derive a reconstruction of the NH
temperature field back to A.D. 1400. This reconstruction was shown to
compare well with the \citet{MaBrHu98} CFR\ldots\ \citet{MaRuWaAm05} attempted
to test the \citet{Ruthetal05} RegEM method using pseudoproxies derived
from the National Center for Atmospheric Research (NCAR) Climate System
Model (CSM) 1.4 millennial integration. Subsequently, \citet{MaRuWaAm07}
have tested a different implementation of RegEM and shown it to perform
favorably in pseudoproxy experiments. This latter study was performed
in part because \citet{MaRuWaAm05} did not actually test the \citet
{Ruthetal05} technique, which was later shown to fail appropriate
pseudoproxy tests [\citet{SmeKap07}]\ldots\ \citet{MaRuWaAm05} used
information during the reconstruction interval, a luxury that is only
possible in the pseudoclimate of a numerical model simulation and not
in actual reconstructions of the earth's climate''.} This is an
especially important point since it is common to find that some methods
work well in some settings and quite poorly in others.\looseness=-1
\item[(3)] SMR make absolutely no attempt to deal with uncertainties, either
for a given draw of data from the simulation or across repeated draws
of the simulation even though there is considerable variation in both
[see \citet{BurCub05} for variation of fit conditional on data and \citet
{ChrSchThe09} for variation of fit across draws of a simulation].
\item[(4)] How relevant are the results of the simulation to the real data
application (i.e., Berliner's point about the ``need to better assess''
these large-scale climate system models, something we return to in
Section \ref{qq} below)?
\end{longlist}

Fortunately, we are able to use the data and code provided to us to
rebut SMR's findings. Before proceeding, however, we must note a
troubling problem with SMR Figure 2. Visual inspection of the plots
reveals an errant feature: OLS methods appear to have nonzero average
residual \textit{in-sample}! Upon examining the code SMR did provide, we
confirmed that this is indeed the case. The culprit is an unreported
and improper centering of the data subsequent to the model fits,
resulting in biased estimates and uncalibrated confidence intervals.

%f3 ###

SMR Figure 2 does not plot raw Northern Hemisphere annual temperature
but rather NH temperature anomaly, that is, NH temperature minus the
average NH temperature over a subset of the in-sample period (defined
to be 1900--1980 AD for the CSM simulation and 1901--1980 for the GKSS
simulation). This centering technique is common in climate science and
simply represents a change of location. However, SMR fit the various
OLS and Lasso methods to the raw (un-centered) temperature over the
full calibration period 1856--1980 AD. In order to center the
predictions, rather than subtracting off the mean 1900--1980 (1901--1980)
AD NH temperature, they subtracted off the mean of \textit{each model's
fitted values} over 1900--1980 (1901--1980) AD. This erroneous and
nonstandard centering results in a substantially biased predictor with
an overestimated RMSE. We refit the models to centered rather than raw
temperature and the RMSEs were about 15--20\% lower than in SMR Table S1
(see SI). Furthermore, the differences between the various methods were
dramatically reduced.

Additionally, SMR make no attempt to grapple with standard errors. As a
first step to address this, we replot SMR Figure 2a appropriately
centered in Figure \ref{fig:smr2a} (for the other three panels, see
SI). In addition to including a corrected version of their smoothed
plot, we also include an unsmoothed plot. As with the real data plotted
in Figure \ref{fig:smr1a}, the differences across methods are dwarfed
by the annual variation within method. Thus, differences among various
methods do not appear so large.

%f4 ###
\begin{figure}%[b]

\includegraphics{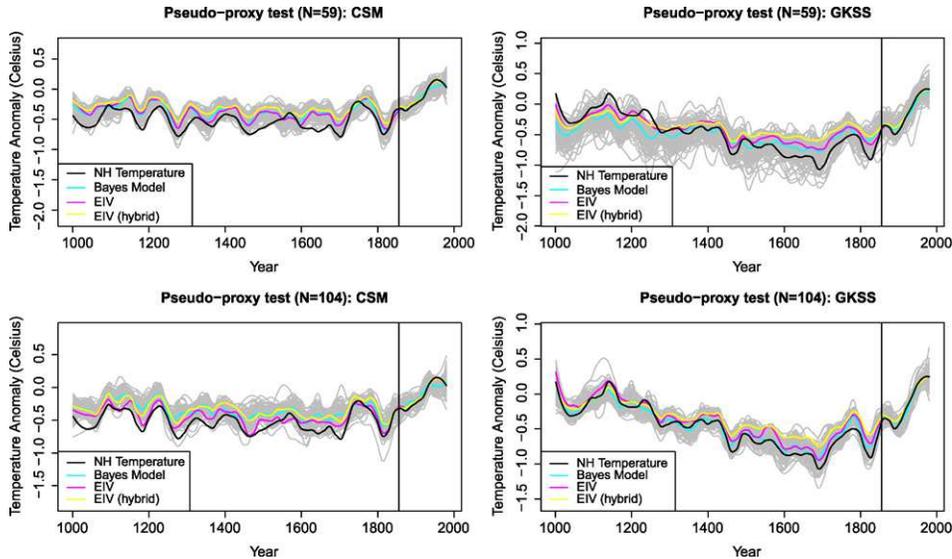}

\caption{Bayesian $\mathit{AR}2+\mathit{PC}10$ model of Section $5$ applied to simulated
data and smoothed. As can be seen, the Bayes model appears to perform
similarly to both RegEM EIV methods. Furthermore, the confidence
intervals of the Bayes model (gray) appear to be calibrated. We
give the unsmoothed version of this figure in the SI.}\label{fig:simbayes10}
\end{figure}

We can improve on SMR Figure 2 and our own Figure \ref{fig:smr2a}
substantially by drawing on our Bayesian model of Section 5. To that
end, we fit the Bayesian $\mathrm{AR}2+\mathrm{PC}10$ model to the simulated data provided
by SMR and include smoothed sample posterior prediction paths in Figure
\ref{fig:simbayes10} (we include an unsmoothed plot in the SI; since
SMR prefer four principal components, we include plots for a Bayesian
$\mathrm{AR}2+\mathrm{PC}4$ model in the SI and note that the four and ten PC models
perform almost identically). As can be seen, our model provides a
prediction which is almost identical to that of two RegEM fits. There
are no statistically significant or even practically important
differences between our model's reconstruction and that of RegEM.
Furthermore, the confidence bands provided by the model generally
include the target NH temperature series (and always do when
unsmoothed), thus suggesting the large uncertainty bands of our Section~\ref{sec5}
are appropriate and not too wide.

In addition, our Bayesian models outperform RegEM EIV in terms of
holdout RMSE (see SI). In fact, they even outperform the hybrid version
of RegEM EIV in two of the four simulations. In a sense, this is not
even a fair comparison because the hybrid method makes use of annual as
well as smoothed (lowpassed) proxy data. Indeed, it is possible to make
a hybrid version of any method, including our Bayesian method, and such
hybrids would be expected to perform better than the nonhybrid version
shown here. However, ``in practice, whether or not the hybrid procedure
is used appears to lead to only very modest differences in skill'' [\citet
{MaRuWaAm07}, page 3].

A final point worth noting is that this demonstration accounts only for
the uncertainty of the model fit conditional on one draw of the
simulation. As stated before, we are unable to properly assess how
model fits vary from draw to draw. This unaccounted for source of
variation is likely large [\citet{ChrSchThe09}] and would be a useful
subject for additional research.

%s1.4 ###
\subsection{Real data versus simulated data}\label{qq}

Berliner calls for an assessment of whether large-scale climate models
like those studied in Section \ref{smr:sim} can serve as a surrogate
for controlled experiments. In this section, we make a modest advance
on this front (see SI for all plots).

Climate scientists, when evaluating these simulations, have focused on
several technical issues. \citet{SmeGonZor08} shows that \citet
{MaRuWaAm07} employed an inappropriate interpolation of GKSS
temperatures and that verification statistics ``are weakened when an
appropriate interpolation scheme is adopted''. More recently, \citet
{SmeKapAmr10} ``identified problems with publicly available versions of
model fields used in \citet{MaRuWaAm05} and \citet{MaRuWaAm07}'' thereby
showing that ``the quantitative results of all pseudoproxy experiments
based on these fields are either invalidated or require
reinterpretation''. Hence, climate scientists have questioned the value
of results derived from the CSM and GKSS simulations due to technical
issues internal to the simulation procedure.

As statisticians, we approach the evaluation of simulated data from a
somewhat different perspective. When statisticians design simulations,
we tend to follow one very important general rubric: if one wants
insights gleaned from the simulation to carry over to real data, then
key features of the simulated data should match key features of the
real data. Thus, we augment climate scientists' ``internal'' evaluation
of the simulated data with an ``external'' evaluation comparing it to
real data.

We have already seen one way in which the real data and simulated data
appear to differ: RegEM gives fits and predictions that are nearly
identical to those of OLS PC4 and OLS G5 PC5 on the real proxy data
(see SMR Figure 1a and our Figure~\ref{fig:smr1a}) but the fits and
predictions on the simulated data are quite different (see SMR Figure 2
and our corrected versions in Figure \ref{fig:smr2a} and the SI; see
also SMR Table S1 and our corrected version given in the SI). More
broadly, we observe that the simulations appear to have smoother NH
temperatures than the real data. This is confirmed by the much smoother
decay in the autocorrelation function. Furthermore, the partial
autocorrelations appear to die out by lag two for the simulations
whereas for the real data they extend to lag four. Moreover, it seems
the simulated time series have only one or at most two distinct
segments, unlike the ``three or possibly four segments'' in CRU
discerned by DL.

In addition to examining the NH temperatures series, we subject the
local grid temperature and (pseudo-)proxy series to a number of
rigorous tests. We examine QQ plots of several summary statistics of
the various local temperature and proxy series with null distributions
provided by the bootstrap [\citet{EfrTib94}]. The first statistic we
consider is the lag one correlation coefficient (see SI as well as
Figure 7 of the paper). We also consider the correlation of each series
with the relevant Northern Hemisphere temperature, calculated over the
instrumental period. Finally, we standardized each series and looked at
the sample standard deviation of the first difference of the
standardized series. In each case, QQ plots reveal that the real data
distributions are strikingly different than those of the simulated
data. In particular, the result about the lag one correlation
coefficient confirms \citeauthor{SmeKap07}'s (\citeyear{SmeKap07}) observation that the ``colored
noise models used in pseudoproxy experiments may not fully mimic the
nonlinear, multivariate, nonstationary characteristics of noise in many
proxy series''.

Important and obvious features of the real data are not replicated in
the simulated data. This therefore puts the results of Section \ref
{smr:sim} (as well as other studies using these simulations for these
purposes) into perspective. How applicable are these results to the
real data when such prominent features fail to match? We can think of
few more fertile areas for future investigation.

%s2 ###
\section{Section 3 revisited}\label{null}

%s2.1 ###
\subsection{The elusiveness of statistical significance}\label{sec2.1}

Section 3 of our paper deals with the statistical significance of
proxy-based reconstructions and how every assessment of statistical
significance depends on the formulation of both the null and
alternative hypotheses. The question is whether proxies can predict
annual temperature in the instrumental period, with significant
accuracy, over relatively short holdout blocks (e.g., 30 years). There
are two main variables: (i) the method used for fitting the data and
(ii) the set of comparison ``null'' benchmarks. In Figures~9 and 10 of
the paper, we chose a single fitting method (the Lasso) and provided
evidence that the choice of benchmark dramatically alters conclusions
about statistical significance. The proxies seem to have some
statistical significance when compared to white noise and weak AR1 null
benchmarks (particularly on front and back holdout blocks) but not
against more sophisticated AR1(Empirical) and Brownian motion null
benchmarks. McIntyre and McKitrick (MM) seem to most clearly understand the purpose of this
section, and we again recognize their contribution for first pointing
out these facts [\citeauthor{McIMcK05b} (\citeyear{McIMcK05b,McIMcK05d})].

Before responding to specific objections raised against the choices we
made to generate Figures 9 and 10 of the paper, we respond to a deep
and statistically savvy point raised by Smerdon: our Lasso-based test
could be ``subject to Type II errors and is unsuitable for measuring
the degree to which the proxies predict temperature''. He suggests that
composite-plus-scale (CPS; see SI for a description) methods might
yield a different result.

%f5 ###
\begin{figure}%[b]

\includegraphics{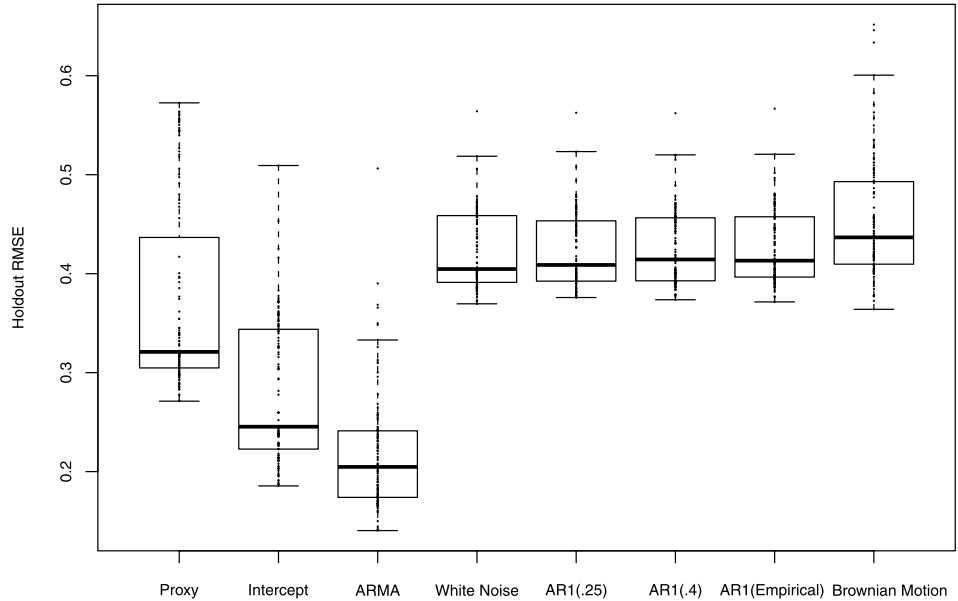}

\caption{Cross-validated RMSE on 30-year holdout blocks for various
models fit to proxies and pseudo-proxies. The procedures used to
generate Intercept and ARMA boxplots are discussed in Section $3.2$ of
the paper. The procedures used to generate the White Noise, $\mathit{AR}1$, and
Brownian motion pseudo-proxies are discussed in Section $3.3$ of the
paper. The CPS fitting procedure used for the Proxy, White Noise,
$\mathit{AR}1$, and Brownian motion boxplots is described in Section
\protect\ref{null} of the long form rejoinder.}\label{fig:cpsboxplot}
\end{figure}

The holdout RMSEs obtained by using CPS on the proxies and
pseudo-proxies (with weights equal to cosine of latitude for Northern
Hemisphere proxies and zero for Southern Hemisphere ones) appear in
Figure \ref{fig:cpsboxplot}. This boxplot has a number of striking
features. Averaged across all holdout blocks, CPS predictions based on
proxies outperform those based on pseudoproxies.\footnote{SMR and
others have chided us for calling our various noise series
``pseudoproxies''. We note, with Tingley, that such series represent the
limiting case of pseudoproxies with zero signal-to-noise ratio and thus
can lay some claim to the name. It is regardless an unimportant
distinction and, for this rejoinder, we stick with the nomenclature of
the paper.} However, CPS performs substantially worse than ARMA models
and certainly no better than an intercept-only model. Finally, the
various pseudoproxy RMSEs are strikingly similar to one another.

In juxtaposition to Figure 9 of the paper which gave the holdout RMSEs
from the Lasso, the CPS holdout RMSEs of Figure \ref{fig:cpsboxplot}
are quite provocative and deserve more attention. Using this
implementation of the CPS method, one might indeed conclude that the
proxies are statistically significant against all pseudoproxies.
However, CPS appears to be a ``weak'' method in that its holdout RMSEs
are larger than the corresponding ones from the Lasso as can be seen by
comparing Figure \ref{fig:cpsboxplot} here to Figure 9 of the paper.
If, in turn, one uses a more powerful method like the Lasso---one that
performs better at the ultimate goal of predicting temperature using
proxies---the results, as indicated in Figure 9 of the paper, are
mixed: the Lasso appears somewhat significant against weak null
benchmarks like AR1(0.25) and AR1(0.4) but not against strong ones like
AR1(Empirical) and Brownian motion.

Actually, the conclusions are decidedly more unclear than the boxplots
of Figure \ref{fig:cpsboxplot} suggest. In the SI, we plot the RMSE by
year for the CPS method and provide null bands based on the sampling
distribution of the pseudo-proxies (as we did for the Lasso in Figure
10 of the paper). We see that the proxies have consistently lower RMSEs
than those of the pseudo-proxies for a majority of the holdout blocks.
Against weak pseudo-proxies, the real proxy predictions are highly
statistically significant on the first few and last few blocks.
However, against more sophisticated pseudo-proxies, CPS forecasts do
not appear to be statistically significant. Furthermore, though much
worse than ARMA models on the interpolation blocks, CPS predictions are
not necessarily worse on the first and last blocks.

Like Aeneas' description of Dido, statistical significance is  \textit{varium et mutabile semper} (fickle and always changing)
[\citet
{Virgil29BC}]. Conditional on a choice of method, pseudoproxy, and
holdout block (or aggregation of blocks), the proxies may appear
statistical significant. Yet, ever so slight variations in those
choices might lead to statistical insignificance. Consequently, it is
easy to misinterpret statistical significance tests and their results
(as also discussed by DL, Kaplan, Smerdon, Tingley, and WA). We fault
many of our predecessors for assiduously collecting and presenting all
the facts that \textit{confirm} their theories while failing to seek facts
that \textit{contradict} them. For science to work properly, it is vital
to stress one's model to its fullest capacity [\citet{Feyn74}]. The
results presented in Figures 9 and 10 of the paper, in Figure \ref
{fig:cpsboxplot} here, and in various figures in the SI, suggest that
maybe our models are in fact not strong enough (or that proxies are too
weak). Furthermore, in contexts where the response series has
substantial ability to locally self-predict, it is vital to recognize
this and make sure the model and covariates provide incremental value
over that [otherwise, ``a particular covariate that is independent of
the response, but is able to mimic the dependence structure of the
response can lead to spurious results'' (DL)]. Methods like the Lasso
coupled with pseudoproxies like AR1(Empirical) and Brownian motion will
naturally account for this self-prediction (see also Kaplan), whereas
naive CPS with latitude-based weighting will not (CPS using univariate
correlation weights does, however; see SI).

%s2.2 ###
\subsection{Specific objections}\label{rmseall}

Specific objections to the results of Section 3 came in two flavors:
(i) criticism of the specific choices we made to get the results
presented in Figures 9 and 10 of the paper, and (ii) questioning the
legitimacy of the entire exercise. The specific criticisms of our
choices were as follows:
\begin{longlist}[(3)]
\item[(1)] The use of the Lasso [Craigmile and Rajaratnam (CR), Haran and
Urban (HU) Rougier, SMR, Tingley, WA]. This is a particularly
interesting criticism since some of our critics (e.g., Kaplan, Rougier)
seem to think the Lasso is a strong method for this context whereas
others (e.g., Tingley, WA) think it weak.
\item[(2)] The use of 30-year holdout blocks (SMR, Smerdon, Tingley).
\item[(3)] The use of interpolated holdout blocks versus extrapolated
holdout blocks (Rougier, Tingley).
\item[(4)] Calibrating our models directly to NH temperature rather than
using local temperatures (Berliner, HU, NL, Tingley).
\end{longlist}
We are able to show, by brute force computation, that our results are
invariant to these choices. Furthermore, as stated in our paper, we
implemented many of these proposals prior to submission (for discussion
of variations originally considered and justification of our choices,
see Section 3.7 for the Lasso; footnote 8 for 30-year blocks; Section
3.4 for interpolation; and Section 3.6 for calibration to local
temperatures). In contrast, we credit  MM for
pointing out the robustness of these results and Kaplan for actually
demonstrating it by using Ridge regression in place of the Lasso (see
Kaplan Figures 1 and 2). We direct the reader to our SI where we
perform the same tests (1) for a plethora of methods (including the
elastic net called for by HU and the Noncentral Lasso called for by
Tingley), (2) using 30- and 60-year holdout blocks, (3) using both
interpolated and extrapolated blocks, and (4) fitting to the local
temperature grid as well as CRU when feasible. Once again, the results
demonstrated by Figures 9 and 10 of our paper are robust to all of
these variations.

The second criticism is more philosophical. WA allege AR1(Empirical)
and Brownian motion pseudoproxies are ``overly conservative'' (a theme
echoed to some extent by HU and SMR) and that ``there is an extensive
literature contradicting \citeauthor{McSWyn11}'s (\citeyear{McSWyn11}) assertions about low or poor
relationships between proxies and climate''. We respond by noting our
pseudoproxies come much closer to mimicking ``the nonlinear,
multivariate, nonstationary characteristics of noise in many proxy
series'' [\citet{SmeKap07}] and by again reflecting on the scope of our
observations. Our paper demonstrates that the relationship between
proxies and temperatures is too weak to detect a rapid rise in
temperatures over short epochs and to accurately reconstruct over a
1000-year period. While there is literature that disagrees with our
conclusions, our explanation is broadly analogous to the statistical
significance results for CPS presented in Figure~\ref{fig:cpsboxplot}:
the relationship between proxies and temperature looks good only for a
weak method and when the self-predictive power of the short NH
temperature sequence (DL) is not properly accounted for.

When it is properly accounted for, statistical insignificance ensues as
demonstrated ably by Kaplan. We therefore endorse Kaplan's assertion
that proxies (whether coupled with ARMA-like models or alone) must
demonstrate statistical significance \textit{above and beyond} ARMA-only
models (Kaplan's ``ability to correct'') and agree with his suggestion
for further research on the matter.

%s3 ###
\section{Two points on Section 4}\label{sec3}

Only two of the discussants (MM and SMR) commented on Section~4 of our
paper. In it, we showed that 27 methods have very similar instrumental
period holdout RMSEs yet provide extremely different temperature
reconstructions [see also \citet{BurCub05}]. This remains true whether
one uses the dataset of 93 proxies from the paper or the dataset of 55
proxies favored by SMR, or whether one uses 30- or 60-year holdout
blocks (see SI; it also appears to broadly hold for data simulated from
climate models as well [\citet{LeeZwiTas08},
\citet{ChrSchThe09}, \citet{SmKaChEv10}]).
Thus, based on predictive ability, one has no reason to prefer ``hockey
sticks'' to ``inverted check marks'' or other shapes and MM are correct
to label this ``a~phenomenon that very much complicates the uncertainty
analysis''.

Also unremarked upon was our point that the proxies seem unable to
capture the high levels of and sharp run-up in temperatures experienced
in the 1990s, even \textit{in-sample} or in contiguous holdout blocks. It
is thus \textit{highly improbable} that they would be able to detect such
high levels and sharp run-ups if they indeed occurred in the more
distant past. That is, we lack statistical evidence that the recently
observed rapid rise in temperature is historically anomalous.

%s4 ###
\section{Section 5 revisited: Bayesian reconstruction}\label{sec4}

We have received a great deal of criticism for our Bayesian
reconstruction of Section 5: for not fully modeling the spatio-temporal
relationships in the data (Berliner, HU, NL, Tingley, SMR), for using a
direct approach rather than an indirect or inverse approach (HU, MM,
NL, Rougier), for linearity (Berliner), and for other features.

The purpose of our model in Section~5 was not to provide a novel
reconstruction method. Indeed, when considering a controversial
question which uses controversial data, new methodologies are likely to
only provoke additional controversy since their properties will be
comparably unknown relative to more tried methods. Rather, we sought a
straightforward model which produces genuine, properly calibrated
posterior intervals and has reasonable out of sample predictive
ability. As Sections~4 and~5 of our paper as well as Section \ref
{smr:sim} show, our model achieves this, providing reconstructive
accuracy as good if not better than the RegEM method as well as
intervals which are properly calibrated. Thus, we take Rougier's
characterization of our model (``perfectly reasonable ad-hockery'') as
high praise. We believe this simple approach is apt, especially for
such a noisy setting. A further virtue of simplicity is that the
model's assumptions are transparent and therefore easy to test and
diagnose. Finally, we believe our model is still among the more
sophisticated models used to produce reconstructions from real proxy
data. Thus far, other more sophisticated approaches have only been
applied to simulated data.

We now turn to the putatively more sophisticated approaches advocated
by our critics [see NL for a very clear exposition; see also \citet
{TinHuy10a} and \citet{LiNycAmm10}]. While these models
have potential advantages, such as a richer spatio-temporal structure,
our experience with real temperature and proxy data causes us to be a
bit more circumspect. These models make a large number of assumptions
about the relationships among global temperature, local temperatures,
proxies, and external forcings. We would like to see a more thorough
investigation of these assumptions because they do not seem to apply to
real data (e.g., how does DL's finding that proxies appear to lead
temperature by 14 years square with such models?). Furthermore, there
are even deeper assumptions embedded in these models which are
difficult to tease out and test on real data.

Hence, we strongly believe that these models need to be rigorously
examined in light of real data. How do they perform in terms of holdout
RMSE and calibration of their posterior intervals? How about when they
are fit to various noise pseudoproxies as in our Section~3? When
replicated data is drawn from the model conditional on the observed
data, does the replicated data  ``look'' like the observed data, especially in
terms of prominent features? In sum, while we believe these models have much to
recommend for themselves and applaud those working on them, we also
strongly believe that tests like those employed in Section~3 of our
paper, Section \ref{qq} of this rejoinder, and various other posterior
predictive checks [\citet{GelCarSte03}, \citet{GelHil06}] are absolutely vital
in the context of such assumption-laden models.

As for the indirect ``multivariate calibration'' approach suggested by
some of the discussants, we point out that it was designed for
highly-controlled almost laboratory-like settings (e.g., chemistry)
with very tight causal relationships. The relationships between
temperature and proxies is considerably dissimilar. Furthermore, we
believe the two approaches, direct and indirect, ought not differ much
in terms of $\hat y$, suggesting that ``both types of procedures should
be able to yield similar results, else we have reason for skepticism''
[\citet{Sun99}]. While one approach or the other might give better
predictions or confidence intervals in this application or that [a~fact that
has been observed even in climate settings; \citet{terBr95}], we believe
Sections~4 and~5 of the paper and Section \ref{smr} of the Rejoinder
suggest our model is adept at both prediction and interval estimation.

%f6 ###
\begin{figure}

\includegraphics{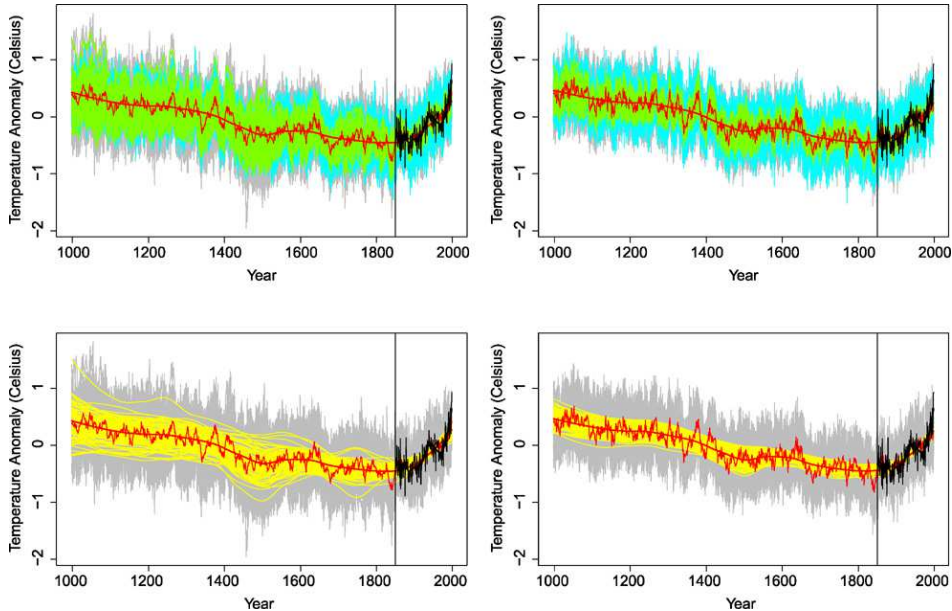}

\caption{Bayesian backcasts and uncertainty decompositions. In the
upper left panel, we re-plot the Bayesian $\mathit{AR}2+\mathit{PC}10$ model from Figure
$16$ of the paper: CRU Northern Hemisphere annual mean land temperature
is given by the thin black line and a smoothed version is given by the
thick black line. The forecast is given by the thin red line and a
smoothed version is given by the thick red line. The model is fit on
1850--1998 AD and backcasts  998--1849 AD. The cyan region
indicates uncertainty due to $\epsilon_t$, the green region indicates
uncertainty due to $\vec\beta$, and the gray region indicates total
uncertainty. In the upper right panel, we give the same plot for a
Bayesian $\mathit{PC}10$ model with no AR coefficients. In the bottom panels, we
re-plot each model's backcast and total uncertainty. We also provide
smooths of each posterior reconstruction path in yellow.}\label{fig:bayessmooth}
\end{figure}

We return to Kaplan's subtle point that the proxies do not necessarily
need to out-predict ARMA-like models, rather that they must simply
provide additional benefits when added to such models. This is a
trenchant point and the dangers of not evaluating proxy reconstructions
in light of ARMA models is illustrated in Figure \ref{fig:bayessmooth}.
The Bayesian $\mathrm{AR}2+\mathrm{PC}10$ model in the upper left and the Bayesian PC10
model in the upper right provide essentially identical reconstructions.
While the PC10 model has a somewhat smaller total posterior interval,
the more striking feature is the disparity in the decomposition. In the
$\mathrm{AR}2+\mathrm{PC}10$ model, most of the uncertainty is due to $\vec\beta$ as
indicated in green; for the PC10 model, the uncertainty due to $\epsilon
_t$ and $\vec\beta$ are more equal in their contribution to total uncertainty.

This has dramatic implications for, among other things, smoothed
reconstructions as shown in the bottom two panels. Smoothing has the
effect of essentially eliminating all uncertainty due to $\epsilon_t$.
Thus, the yellow region in the bottom right plot is extremely narrow
(which explains why confidence intervals in the climate science
literature are typically so narrow; see Figure~17 of the paper). On the
other hand, when an AR2 structure is added, even smoothed confidence
bands are quite wide. This is a profoundly important point which
highlights the necessity of modeling the temporal dependence of the NH
temperature series and we thank Kaplan for raising it.

%s5 ###
\section{Statistical power}\label{sec5}

Our results of Section~3 do not depend on the Lasso and are robust to
changes in the null distribution (i.e., the pseudoproxies), the fitting
algorithm, the holdout period length and location, and the calibration
target series. Nonetheless, there was substantial criticism of the
Lasso by a number of discussants (CR, HU, Rougier, SMR, Tingley, WA)
and worries that our tests lacked statistical power (Smerdon). In this
section, we discuss two of those criticisms (Tingley and Smerdon) and
show that lack of power may be intrinsic to the data at hand.

%s5.1 ###
\subsection{Tingley}\label{tingley}

Tingley asserts that the Lasso ``is simply not an appropriate tool for
reconstructing paleoclimate'' and purports to show this via a simulation
study which has two components. The second of the two components
(featured in the right-hand plots of Tingley Figure 1 and in Tingley
Figure 2) does an exemplary job of showing how the Lasso can use
autocorrelated predictors in order to provide excellent fits of an
autocorrelated response series---even when the response and predictors
are generated independently.\looseness=-1

The first component of the simulation (featured in the left-hand plots
of Tingley Figure 1) is problematic for a number of reasons. First, it
is not clear to us how this simulation relates to proxy-based
reconstruction of temperature. If one compares and contrasts plots of
Tingley's simulated data (see SI) to Figures 5 and 6 of the paper, one
sees that his target ``temperature'' series fails to look like the real
temperature series and his pseudo-proxies fail to look like the real proxies.

Second, Tingley implements the Lasso in a completely nonstandard way:
``The Lasso penalization parameter [$\lambda$ on page 13 of \citet
{McSWyn11}] is set to be $0.05$ times the smallest value of $\lambda$
for which all coefficients are zero''. There is no apparent statistical
justification for this choice, and, when $\lambda$ is selected through
ten repetitions of five-fold cross-validation (as is done throughout
our paper), the Lasso RMSE is twice as good as in Tingley's Figure 1
(see SI).

Third, we must consider the two methods under consideration. This
simulation is \textit{exactly} the kind of situation where the Lasso is
known to perform poorly. When one has identical predictors each with
the same coefficient, ``the Lasso problem breaks down'' and methods like
ridge regression are superior [\citet{ZouHas05}, \citet{FriHasTib2010}].
Furthermore, Tingley's benchmark of composite regression is both
unrealistically good for this simulation and utterly nonrobust
(furthermore, it also fails to reject the null tests of Section 3; see SI).

Composite regression performs unrealistically well because it is a
univariate linear regression of $y_t$ on a series which roughly equals
$y_t + \nu_t$ where $\nu_t \iid N(0, \sigma_\omega/\sqrt{1138})$ (where
$\sigma_\omega$ is set to various levels by Tingley). It is impossible
for any method to perform comparably well against such an ideal
procedure (one that has asymptotic zero RMSE as the number of
pseudoproxies goes to infinity). Ridge regression, known to perform
well in settings like this simulation, does 1--6 times worse than
composite regression (see SI) and is therefore not much better than the
Lasso (in fact, it is worse for the high noise settings). Even the true
data-generating model for $y_t$ \textit{with the true
parameters}---another unrealistically strong model---performs about 13
times worse than composite regression in some settings (see SI).

%f7 ###
\begin{figure}

\includegraphics{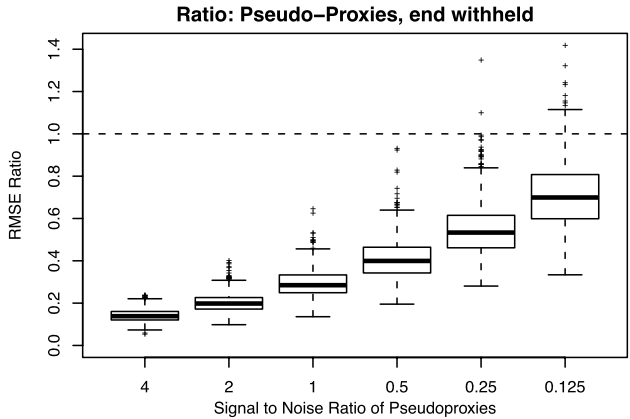}

\caption{Tingley simulation perturbed with $\sigma_\beta=3$. We plot
the ratio of the RMSE of the Lasso to that of composite regression.
Compare to the lefthand plots of Tingley Figure $1$. In the SI, we give
the raw RMSEs of two methods as well as their ratio for this value of
$\sigma_\beta$ and others.}\label{fig:tingleyperturbsigma3}
\end{figure}

Additionally, composite regression lacks robustness to slight but
realistic perturbations of the simulation. For instance, consider
setting $x_{t,i} = \beta_i y_{t} + \omega_{t,i}$ where $\beta_i \iid
N(1,\sigma_\beta=3)$ (Tingley's simulation corresponds to $\sigma_\beta
=0$). RMSE boxplots for this simulation appear in Figure \ref
{fig:tingleyperturbsigma3}. In this case, the Lasso dominates composite
regression by more than a factor of five in the lowest noise case. In
fact, composite regression appears to do arbitrarily badly relative to
the Lasso for high values of $\sigma_\beta$ (see SI for $\sigma_\beta
=1/3,1,9,$ and $27$ in addition to the $\sigma_\beta=3$ presented here).

Thus, it is not the Lasso but the simulation that is broken: the Lasso
is used in a setting where it is known to perform poorly, is implemented in a non-standard fashion, and is pitted
against an unrealistically good and nonrobust competitor model.
Furthermore, it is unclear how this simulation relates to proxy-based
temperature reconstruction.

Tingley's simulation does, however, raise a subtle issue. He shows the
Lasso, when fit to strong AR1 and Brownian motion pseudoproxies that
contain no signal, can provide better predictions than when fit to some
of his pseudoproxies which do contain signal. Taken together, this
suggests that we may never find statistical significance when given
many weakly informative proxy series and thus we lack power.\footnote
{Though it is beyond the scope of our work, we note that, by making use
of additional information (e.g., the spatial locations of the proxies
and local temperatures), it is possible that the proxies might become
considerably more predictive/informative than they have so far proven
to be.} We argue that if this is the case (as it indeed may be; see our
discussion of Smerdon), it is something endemic to all methods, even
his composite regression and the Noncentral Lasso (see~SI).

As a final point, Tingley claims it ``is simply not the case'' that
dimensionality reduction is necessary for the paleoclimate
reconstruction problem, and he suggests Bayesian hierarchical models as
``a more scientifically sound approach''. This is an odd comment since
Bayesian hierarchical models are well known to reduce dimensionality in
the parameter space via partial pooling [\citet{GelCarSte03}, \citet{GelHil06}].
We thus reiterate our claim that dimensionality reduction is intrinsic
to the endeavor.

%s5.2 ###
\subsection{Smerdon}\label{smerdon}

Smerdon suggests a highly sophisticated test of whether or not the
Lasso has power in this paleoclimate context. His simulation technique
is to increasingly corrupt local temperatures and compare how the Lasso
performs on these series to the proxy and pseudoproxy series. This is
quite clever since, by definition, local temperatures contain signal
for NH temperature.

\begin{figure}[t]

\includegraphics{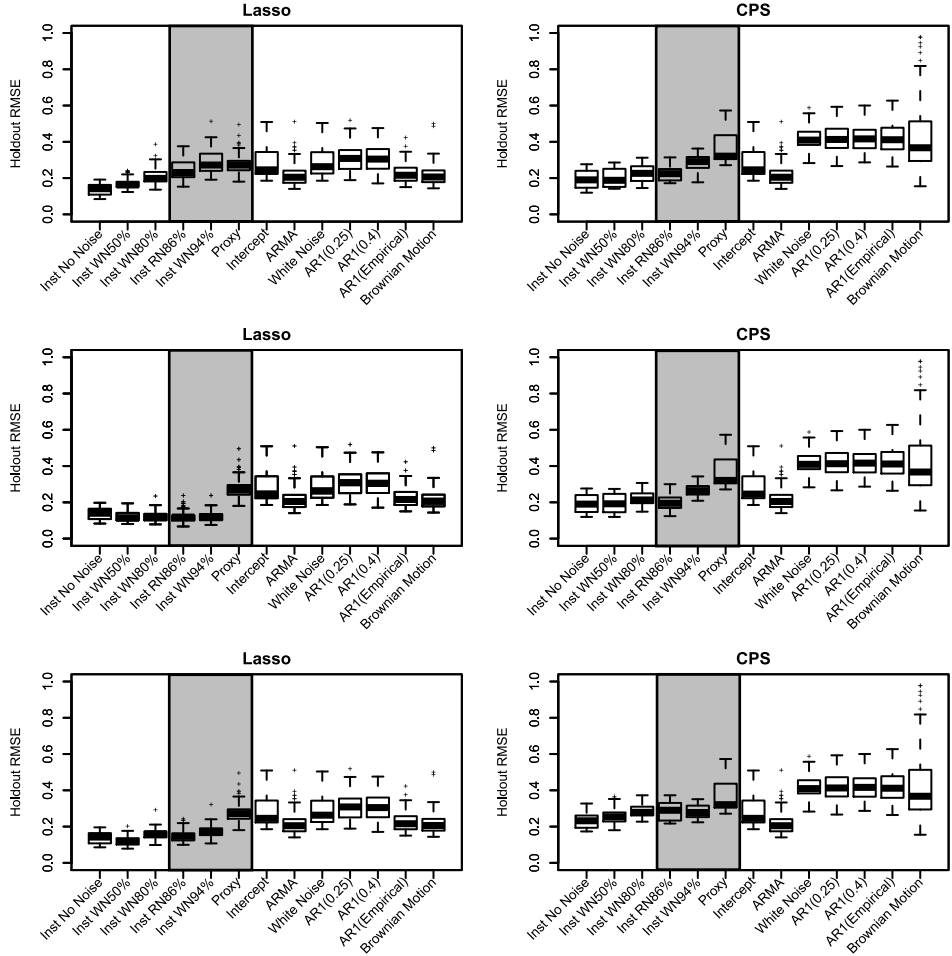}

\caption{Comparison of Lasso and CPS on various pseudoproxies. In the
top left, we plot give holdout RMSEs for the Lasso trained using
Smerdon's instrumental period local temperature pseudoproxies (this
figure replicates Smerdon Figure 1a). In the top right, we give the
same plot for CPS. In the middle row and bottom rows, we consider two
variants of Smerdon's pseudoproxies (see text for more detail).
Compare to Smerdon Figure 1a and the SI.}\label{fig:smerdon}
\end{figure}

The first thing of note that Smerdon shows is that ``even `perfect
proxies' are subject to errors'' (see Smerdon Figure 1a which is reproduced as the
top left panel of Figure \ref{fig:smerdon}), a fact we have noticed in
our own work (see SI). Local instrumental temperatures have substantial
error when predicting CRU NH temperature on holdout blocks.

Smerdon also shows that the proxies perform similarly to local
temperatures corrupted with either 86\% red noise or 94\% white noise
(see the top left panel of Figure \ref{fig:smerdon} which reproduces
Smerdon Figure 1a). On the other hand, our AR1(Empirical) and Brownian
motion pseudoproxies outperform the corrupted temperatures suggesting
that our test rejects even ``proxies'' known to have signal (albeit a
highly corrupted one). We agree with Smerdon that these results come
with a number of caveats,\footnote{Smerdon samples the temperature grid
only once and he samples from the whole globe as opposed to either
sampling from the NH or using the locations of the \citet{Mannetal08}
proxies. He also samples the noise series only once for each setting.
Furthermore, he conducts only one repetition for each holdout block.
Finally, he compares the Lasso performance on 283 predictors
constructed from local temperatures to performance on 1138 proxies and
noise pseudoproxies, thus lowering $p$ substantially. We believe
consideration of these factors are unlikely to alter the basic picture
presented in the top left panel of Figure \ref{fig:smerdon}. However,
it would likely increase the variance of the various boxplots thus
making the differences less stark. Moreover, it would be interesting to
see the RMSE distributions from holdout block to holdout block along
with intervals for resampled temperatures and noise pseudoproxies
(i.e., in the style of Figure 10 versus Figure 9 of the paper).} but we
believe they warrant more reflection.

%f8 ###

Smerdon says ``skillful CPS reconstructions (latitude-based weights)
can be derived from such predictors'' (i.e., from corrupted
temperatures) and presents Smerdon Figure 1b as evidence. This figure
is problematic and misleading for a number of reasons. First, it is
entirely in-sample. Second, it omits the Lasso's performance. In fact,
the Lasso gives a lower RMSE at the same task (CPS has an RMSE of 0.223
and 0.266 on the 86\% red noise and 94\% white noise corrupted
temperatures respectively whereas the Lasso has 0.070 and 0.108, respectively).

A fairer comparison would be to use the test of Smerdon Figure 1a on
CPS, which we present in the top right panel of Figure \ref
{fig:smerdon}.\footnote{The eight right-most boxplots of the top panel
of Figure \ref{fig:smerdon} should be reminiscent of Figure \ref
{fig:cpsboxplot}. They are identical except the former is based on one
repetition for each holdout block whereas the latter averages over 100
such repetitions thus decreasing variation.} As can be seen, CPS has
lower RMSEs on the corrupted instrumental temperatures than on noise
pseudoproxies such as AR1(Empirical) and Brownian motion. The proxies
also appear to perform better than these noise pseudoproxies. So it
seems that CPS passes Smerdon's test. On the other hand, the Lasso
performs equivalently to or better than CPS in all 13 cases. Hence, it
does not seem that CPS reconstructions are so skillful after all since
they are worse than those of the Lasso.

As mentioned in the response to Tingley, the Lasso is known to perform
poorly in situations where all the predictors are the same. This is
approximately the situation governing the five leftmost boxplots in the
two upper panels of Figure \ref{fig:smerdon}. In contrast, we consider
two variants of Smerdon's simulation. The first variant, plotted in the
second row of Figure \ref{fig:smerdon}, defines noise percentage
differently. Rather than adding noise to local temperatures such that
the variance of the noise accounts for some fixed percentage of the
variance of the sum, we add predictors to the matrix of local
temperatures such that a fixed percentage of the predictors are pure
noise (e.g., for 50\% white noise, the matrix of predictors has 566
columns, 283 of local temperatures and 283 of white noise). The Lasso
performs spectacularly well in this setting and seems fully powered
with the local temperatures always outperforming various noise series.
On the other hand, CPS performs similarly in this case as in Smerdon's
example: local temperatures outperform noise series but the predictions
on the whole are quite poor.

The second variant of Smerdon's simulation replicates what we did with
Tingley's simulation. Rather than using local temperature plus noise as
predictors, we used a random slope times local temperature plus noise
where the slopes were i.i.d. $N(1,\sigma_\beta=3)$. In some sense, this
better reflects the relationship between proxies and temperature. In
this setting, the Lasso again performs very strongly with the corrupted
local temperatures always outperforming various noise series. CPS gives
worse results across the board. For results using different values of
$\sigma_\beta$, see SI.

In sum, when predictors have approximately the same coefficient \textit{and} there is a very high noise level
(e.g., the 86\% and 94\% noise
conditions of the top left panel of Figure \ref{fig:smerdon}), the
Lasso is perhaps underpowered. In variants of the simulation that might
be more true to real data (the middle left and bottom left panels), the
Lasso performs very well. On the other hand, CPS performs weakly in all
settings: it simply does not provide particularly good out of sample
predictions compared to methods like the Lasso.

We are left wondering why the Lasso ``fails'' Smerdon's test, suggesting
a lack of power. Low power can be explained by recognizing an
unavoidable reality: the NH temperature sequence is short, highly
autocorrelated, and ``blocky'' (DL observe ``3 or possibly 4 segments'').
Thus, the effective sample size is far smaller than $n=149$.
Consequently, as is always the case with small samples, one lacks
power. It therefore follows (as shown in Section \ref{rmseall} and the
figures in the Appendix to the SI) that failure to reject the null
against AR1(Empirical) and Brownian motion pseudo-proxies is not
specific to the Lasso. Rather, it is endemic to the problem. Unless we
can find proxies that strongly predict temperature at an annual level,
power will necessarily be low and uncertainty high.

%s6 ###
\section{Conclusion}\label{sec6}

In conclusion, we agree with Berliner that statisticians should ``not
continue with questionable assumptions, nor merely offer small fixes to
previous approaches, nor participate in uncritical debates'' of climate
scientists. Nonetheless, we believe that these assumptions (linearity,
stationarity, data quality, etc.) were clearly stated in the beginning
of our paper and were not endorsed. We believe it was important to
start with these questionable, perhaps even indefensible, assumptions
to engage the literature to date and to provide a point of departure
for future work.

We also reiterate our conclusion that ``climate scientists have greatly
underestimated the uncertainty of proxy-based reconstructions and hence
have been overconfident in their models''. In fact, there is reason to
believe the wide confidence intervals given by our model of Section 5
are optimistically narrow. First, while we account for parameter
uncertainty, we do not take model uncertainty into account. Second, we
take the data as given and do not account for uncertainties, errors,
and biases in selection, processing, in-filling, and smoothing of the
data as well as the possibility that the data has been ``snooped''
(subconsciously or otherwise) based on key features of the first and
last block. Since these features are so well known, there is absolutely
no way to create a dataset in a ``blind'' fashion. Finally, and perhaps
most importantly, the NRC assumptions of linearity and stationarity
[\citet{NRC06}] outlined in our paper are likely untenable and we agree
with Berliner in calling them into question. While the ``infinite
confidence intervals'' of the \citet{BroSun87} test reported by MM are
unrealistically large due to physical constraints, we agree with their
central point that this matter warrants closer examination since it is
absolutely critical to assessing statistical significance and
predictive accuracy.

As a final point, numerous directions for future research appear in
this paper, discussion, and rejoinder. We hope statisticians and
climate scientists will heed the call on these problems.

\section*{Acknowledgments}
We thank Jason Smerdon for a number of helpful conversations about
data, methods, and code. Those conversations were tremendously fecund
and greatly enhanced our rejoinder. We believe such conversations to be
paradigmatic of the great value of collaboration between climate
scientists and statisticians and hope they can serve as a template for
future work between the two camps. We also thank Steve McIntyre for
several very helpful discussions about data and code.

\begin{supplement}[id=suppA]
\sname{Supplement A}
\stitle{Long form rejoinder:
A statistical analysis of multiple temperature proxies: Are
reconstructions of surface temperatures over the last
1000 years reliable?}
\slink[doi]{10.1214/10-AOAS398REJSUPPA} %[doi,text={...}] - jei reikia
%suskaldyti doi
\slink[url]{http://lib.stat.cmu.edu/aoas/398REJ/supplementREJ_A.zip}
\sdatatype{.zip}
\sdescription{This document is the long form of
``Rejoinder: A statistical analysis of multiple temperature proxies:
Are reconstructions of surface temperatures
over the last 1000 years reliable?'' It contains all the text from the
short form which appeared in print as well
as the supporting details and figures for the claims made in that document.}
\end{supplement}
\begin{supplement}[id=suppB]
\sname{Supplement B}
\stitle{Code repository for ``Rejoinder: A statistical analysis of
multiple temperature proxies: Are
reconstructions of surface temperatures over the last 1000 years reliable?''}
\slink[doi]{10.1214/10-AOAS398REJSUPPB} %[doi,text={...}] - jei reikia
%suskaldyti doi
\slink[url]{http://lib.stat.cmu.edu/aoas/398REJ/supplementREJ_B.zip}
\sdatatype{.zip}
\sdescription{This repository archives all data and code used for
``Rejoinder: A statistical analysis of multiple temperature proxies: Are
reconstructions of surface temperatures over the last
1000 years reliable?''
In particular, it contains code to make all figures and tables featured
in the long form
(which is a superset of those in the short form).}
\end{supplement}

% imsref loaded by smiklovaite, 2011-01-20 08:45:01
% imsref loaded by smiklovaite, 2011-01-20 08:50:03
% imsref loaded by smiklovaite, 2011-01-20 10:21:55
%

\printaddresses


\begin{thebibliography}{38}
% BibTex style file: ims.bst, 2010-01-14
% Default style options (sort=0,type=number).
% Used options (sort=1,type=nameyear).

%b1 ###
\bibitem[\protect\citeauthoryear{Brown and Sundberg}{1987}]{BroSun87}
%
\begin{barticle}[mr]
\bauthor{\bsnm{Brown},~\bfnm{Philip~J.}\binits{P.~J.}} \AND
\bauthor{\bsnm{Sundberg},~\bfnm{Rolf}\binits{R.}}
(\byear{1987}).
\btitle{Confidence and conflict in multivariate calibration}.
\bjournal{J.~Roy. Statist. Soc. Ser. B}
\bvolume{49}
\bpages{46--57}.
\bid{mr={0893336}}
\end{barticle}
%
\endbibitem

%b2 ###
\bibitem[\protect\citeauthoryear{Burger and Cubasch}{2005}]{BurCub05}
%
\begin{barticle}[auto:STB|2010-11-18|09:18:59]
\bauthor{\bsnm{Burger},~\bfnm{G.}\binits{G.}} \AND
\bauthor{\bsnm{Cubasch},~\bfnm{U.}\binits{U.}}
(\byear{2005}).
\btitle{Are multiproxy climate reconstructions robust?}
\bjournal{Geophysical Research Letters}
\bvolume{32}
\bpages{L23711}.
\end{barticle}
%
\endbibitem

%b3 ###
\bibitem[\protect\citeauthoryear{Christiansen, Schmith and
Thejll}{2009}]{ChrSchThe09}
%
\begin{barticle}[auto:STB|2010-11-18|09:18:59]
\bauthor{\bsnm{Christiansen},~\bfnm{B.}\binits{B.}},
\bauthor{\bsnm{Schmith},~\bfnm{T.}\binits{T.}} \AND
\bauthor{\bsnm{Thejll},~\bfnm{P.}\binits{P.}}
(\byear{2009}).
\btitle{A surrogate ensemble study of climate reconstruction methods:
Stochasticity and robustness}.
\bjournal{Journal of Climate}
\bvolume{22}
\bpages{951--976}.
\end{barticle}
%
\endbibitem

%b4 ###
\bibitem[\protect\citeauthoryear{Christiansen, Schmith and
Thejll}{2010}]{ChrSchThe10}
%
\begin{barticle}[auto:STB|2010-11-18|09:18:59]
\bauthor{\bsnm{Christiansen},~\bfnm{B.}\binits{B.}},
\bauthor{\bsnm{Schmith},~\bfnm{T.}\binits{T.}} \AND
\bauthor{\bsnm{Thejll},~\bfnm{P.}\binits{P.}}
(\byear{2010}).
\btitle{Reply}.
\bjournal{Journal of Climate}
\bvolume{23}
\bpages{2839--2844}.
\end{barticle}
%
\endbibitem

%b5 ###
\bibitem[\protect\citeauthoryear{Diaconis}{1985}]{Diac85}
%
\begin{bincollection}[auto:STB|2010-11-18|09:18:59]
\bauthor{\bsnm{Diaconis},~\bfnm{P.}\binits{P.}}
(\byear{1985}).
\btitle{Theories of data analysis: From magical thinking through classical
statistics}.
In \bbooktitle{Exploring Data Tables, Trends and Shapes}
\bpages{1--36}.
\bpublisher{Wiley}, \baddress{New York}.
\end{bincollection}
%
\endbibitem

%b6 ###
\bibitem[\protect\citeauthoryear{Efron and Tibshirani}{1994}]{EfrTib94}
%
\begin{bbook}[mr]
\bauthor{\bsnm{Efron},~\bfnm{Bradley}\binits{B.}} \AND
\bauthor{\bsnm{Tibshirani},~\bfnm{Robert~J.}\binits{R.~J.}}
(\byear{1994}).
\btitle{An Introduction to the Bootstrap}.
\bseries{Monographs on Statistics and Applied Probability}
\bvolume{57}.
\bpublisher{Chapman \& Hall}, \baddress{New York}.
\end{bbook}
%
\endbibitem

%b7 ###
\bibitem[\protect\citeauthoryear{Feynman}{1974}]{Feyn74}
%
\begin{barticle}[auto:STB|2010-11-18|09:18:59]
\bauthor{\bsnm{Feynman},~\bfnm{R.}\binits{R.}}
(\byear{1974}).
\btitle{Cargo cult science}.
\bjournal{Engineering and Science}
\bvolume{37}
\bpages{7}.
\end{barticle}
%
\endbibitem

%b8 ###
\bibitem[\protect\citeauthoryear{Friedman, Hastie and
Tibshirani}{2010}]{FriHasTib2010}
%
\begin{barticle}[pbm]
\bauthor{\bsnm{Friedman},~\bfnm{Jerome}\binits{J.}},
\bauthor{\bsnm{Hastie},~\bfnm{Trevor}\binits{T.}} \AND
\bauthor{\bsnm{Tibshirani},~\bfnm{Rob}\binits{R.}}
(\byear{2010}).
\btitle{Regularization paths for generalized linear models via coordinate
descent}.
\bjournal{J. Stat. Softw.}
\bvolume{33}
\bpages{1--22}.
\bid{pmid={20808728}, pmcid={2929880}, mid={NIHMS201118}}
\end{barticle}
\endbibitem

%b9 ###
\bibitem[\protect\citeauthoryear{Gelman and Hill}{2006}]{GelHil06}

\begin{bbook}[auto:STB|2010-11-18|09:18:59]
\bauthor{\bsnm{Gelman},~\bfnm{A.}\binits{A.}} \AND
\bauthor{\bsnm{Hill},~\bfnm{J.}\binits{J.}}
(\byear{2006}).
\btitle{Data Analysis Using Regression and Multilevel/Hierarchical Models}.
\bpublisher{Cambridge Univ. Press}, \baddress{New York}.
\end{bbook}
\endbibitem

%b10 ###
\bibitem[\protect\citeauthoryear{Gelman et al.}{2003}]{GelCarSte03}
%
\begin{bbook}[mr]
\bauthor{\bsnm{Gelman},~\bfnm{Andrew}\binits{A.}},
\bauthor{\bsnm{Carlin},~\bfnm{John~B.}\binits{J.~B.}},
\bauthor{\bsnm{Stern},~\bfnm{Hal~S.}\binits{H.~S.}} \AND
\bauthor{\bsnm{Rubin},~\bfnm{Donald~B.}\binits{D.~B.}}
(\byear{2003}).
\btitle{Bayesian Data Analysis},
\bedition{2nd} ed.
\bpublisher{Chapman \& Hall}, \baddress{Boca Raton, FL}.
\end{bbook}
\endbibitem

%b11 ###
\bibitem[\protect\citeauthoryear{Hastie, Tibshirani and
Friedman}{2009}]{HasTibFri09}
\begin{bbook}[mr]
\bauthor{\bsnm{Hastie},~\bfnm{Trevor}\binits{T.}},
\bauthor{\bsnm{Tibshirani},~\bfnm{Robert}\binits{R.}} \AND
\bauthor{\bsnm{Friedman},~\bfnm{Jerome}\binits{J.}}
(\byear{2009}).
\btitle{The Elements of Statistical Learning: Data Mining, Inference,
and Prediction},
\bedition{2nd} ed.
\bpublisher{Springer}, \baddress{New York}.
\bid{doi={10.1007/978-0-387-84858-7}, mr={2722294}}
\end{bbook}
\endbibitem

%b12 ###
\bibitem[\protect\citeauthoryear{Izenman}{2008}]{Izen08}
\begin{bbook}[mr]
\bauthor{\bsnm{Izenman},~\bfnm{Alan~Julian}\binits{A.~J.}}
(\byear{2008}).
\btitle{Modern Multivariate Statistical Techniques: Regression,
Classification, and Manifold Learning}.
\bpublisher{Springer}, \baddress{New York}.
\bid{doi={10.1007/978-0-387-78189-1}, mr={2445017}}
\end{bbook}
\endbibitem

%b13 ###
\bibitem[\protect\citeauthoryear{Lee, Zwiers and Tsao}{2008}]{LeeZwiTas08}
%
\begin{barticle}[auto:STB|2010-11-18|09:18:59]
\bauthor{\bsnm{Lee},~\bfnm{T.~C.~K.}\binits{T.~C.~K.}},
\bauthor{\bsnm{Zwiers},~\bfnm{F.~W.}\binits{F.~W.}} \AND
\bauthor{\bsnm{Tsao},~\bfnm{M.}\binits{M.}}
(\byear{2008}).
\btitle{Evaluation of proxy-based millennial reconstruction methods}.
\bjournal{Climate Dynamics}
\bvolume{31}
\bpages{263--281}.
\end{barticle}
%
\endbibitem

%b14 ###
\bibitem[\protect\citeauthoryear{Li, Nychka and Amman}{2010}]{LiNycAmm10}
%
\begin{barticle}[auto:STB|2010-11-18|09:18:59]
\bauthor{\bsnm{Li},~\bfnm{B.}\binits{B.}},
\bauthor{\bsnm{Nychka},~\bfnm{D.~W.}\binits{D.~W.}} \AND
\bauthor{\bsnm{Amman},~\bfnm{C.~M.}\binits{C.~M.}}
(\byear{2010}).
\btitle{The value of multi-proxy reconstructin of past climate}.
\bjournal{J. Amer. Statist. Assoc.}
\bvolume{105}
\bpages{883--895}.
\end{barticle}
%
\endbibitem

%b15 ###
\bibitem[\protect\citeauthoryear{Little and Rubin}{2002}]{LitRub02}
%
\begin{bbook}[mr]
\bauthor{\bsnm{Little},~\bfnm{Roderick J.~A.}\binits{R.~J.~A.}} \AND
\bauthor{\bsnm{Rubin},~\bfnm{Donald~B.}\binits{D.~B.}}
(\byear{2002}).
\btitle{Statistical Analysis with Missing Data},
\bedition{2nd} ed.
\bpublisher{Wiley}, \baddress
{Hoboken, NJ}.
\bid{mr={1925014}}
\end{bbook}
%
\endbibitem

%b16 ###
\bibitem[\protect\citeauthoryear{Mann, Bradley and Hughes}{1998}]{MaBrHu98}
%
\begin{barticle}[auto:STB|2010-11-18|09:18:59]
\bauthor{\bsnm{Mann},~\bfnm{M.~E.}\binits{M.~E.}},
\bauthor{\bsnm{Bradley},~\bfnm{R.~E.}\binits{R.~E.}} \AND
\bauthor{\bsnm{Hughes},~\bfnm{M.~K.}\binits{M.~K.}}
(\byear{1998}).
\btitle{Global-scale temperature patterns and climate forcing over the
past six
centuries}.
\bjournal{Nature}
\bvolume{392}
\bpages{779--787}.
\end{barticle}
%
\endbibitem

%b17 ###
\bibitem[\protect\citeauthoryear{Mann et al.}{2005}]{MaRuWaAm05}
%
\begin{barticle}[auto:STB|2010-11-18|09:18:59]
\bauthor{\bsnm{Mann},~\bfnm{M.~E.}\binits{M.~E.}},
\bauthor{\bsnm{Rutherford},~\bfnm{S.}\binits{S.}},
\bauthor{\bsnm{Wahl},~\bfnm{E.}\binits{E.}} \AND
\bauthor{\bsnm{Ammann},~\bfnm{C.}\binits{C.}}
(\byear{2005}).
\btitle{Testing the fidelity of methods used in proxy-based
reconstructions of
past climate}.
\bjournal{Journal of Climate}
\bvolume{18}
\bpages{4097--4107}.
\end{barticle}
%
\endbibitem

%b18 ###
\bibitem[\protect\citeauthoryear{Mann et al.}{2007}]{MaRuWaAm07}
%
\begin{barticle}[auto:STB|2010-11-18|09:18:59]
\bauthor{\bsnm{Mann},~\bfnm{M.~E.}\binits{M.~E.}},
\bauthor{\bsnm{Rutherford},~\bfnm{S.}\binits{S.}},
\bauthor{\bsnm{Wahl},~\bfnm{E.}\binits{E.}} \AND
\bauthor{\bsnm{Ammann},~\bfnm{C.}\binits{C.}}
(\byear{2007}).
\btitle{Robustness of proxy-based climate field reconstruction methods}.
\bjournal{Journal of Geophysical Research}
\bvolume{112}
\bpages{D12109}.
\bid{doi={10.1029/2006JD008272}}
\end{barticle}
%
\endbibitem

%b19 ###
\bibitem[\protect\citeauthoryear{Mann et al.}{2008}]{Mannetal08}
%
\begin{barticle}[auto:STB|2010-11-18|09:18:59]
\bauthor{\bsnm{Mann},~\bfnm{M.~E.}\binits{M.~E.}},
\bauthor{\bsnm{Zhang},~\bfnm{Z.}\binits{Z.}},
\bauthor{\bsnm{Hughes},~\bfnm{M.~K.}\binits{M.~K.}},
\bauthor{\bsnm{Bradley},~\bfnm{R.~S.}\binits{R.~S.}},
\bauthor{\bsnm{Miller},~\bfnm{S.~K.}\binits{S.~K.}},
\bauthor{\bsnm{Rutherford},~\bfnm{S.}\binits{S.}} \AND
\bauthor{\bsnm{Ni},~\bfnm{F.}\binits{F.}}
(\byear{2008}).
\btitle{Proxy-based reconstructions of hemispheric and global surface
temperature variations over the past two millenia}.
\bjournal{Proc. Natl. Acad. Sci. USA}
\bvolume{105}
\bpages{36}.
\end{barticle}
%
\endbibitem

%b20 ###
\bibitem[\protect\citeauthoryear{Maro}{29BC}]{Virgil29BC}
%
\begin{bmisc}[auto:STB|2010-11-18|09:18:59]
\bauthor{\bsnm{Maro},~\bfnm{P.~V.}\binits{P.~V.}}
(\byear{29BC}).
\bhowpublished{\textit{Aeneid}. Octavius Press, Rome}.
\end{bmisc}
%
\endbibitem

%b21 ###
\bibitem[\protect\citeauthoryear{McIntyre and McKitrick}{2005a}]{McIMcK05b}
%
\begin{barticle}[auto:STB|2010-11-18|09:18:59]
\bauthor{\bsnm{McIntyre},~\bfnm{S.}\binits{S.}} \AND
\bauthor{\bsnm{McKitrick},~\bfnm{R.}\binits{R.}}
(\byear{2005}a).
\btitle{Hockey sticks, principal components, and spurious significance}.
\bjournal{Geophysical Research Letters}
\bvolume{32}
\bpages{L03710}
\bid{doi={10.1029/2004GL021750}}.
\end{barticle}
%
\endbibitem

%b22 ###
\bibitem[\protect\citeauthoryear{McIntyre and McKitrick}{2005b}]{McIMcK05d}
%
\begin{barticle}[auto:STB|2010-11-18|09:18:59]
\bauthor{\bsnm{McIntyre},~\bfnm{S.}\binits{S.}} \AND
\bauthor{\bsnm{McKitrick},~\bfnm{R.}\binits{R.}}
(\byear{2005}b).
\btitle{Reply to comment by von Storch and Zorita on ``Hockey sticks,
principal components, and spurious significance''.}
\bjournal{Geophysical Research Letters}
\bvolume{32}
\bpages{L20714}
\bid{doi={10.1029/2005GL023089}}.
\end{barticle}
%
\endbibitem

%b23 ###
\bibitem[\protect\citeauthoryear{McShane and Wyner}{2011a}]{McSWyn11}
%
\begin{bmisc}[auto:STB|2010-11-18|09:18:59]
\bauthor{\bsnm{McShane},~\bfnm{B.~B.}\binits{B.~B.}} \AND
\bauthor{\bsnm{Wyner},~\bfnm{A.~J.}\binits{A.~J.}}
(\byear{2011}a).
\bhowpublished{A statistical analysis of multiple temperature proxies: Are
reconstructions of surface temperatures over the last 1000 years reliable?
\textit{Ann. Appl. Stat.} \textbf{5} 5--44}.
\end{bmisc}
%
\endbibitem

%b24 ###
\bibitem[\protect\citeauthoryear{McShane and Wyner}{2011b}]{McSWyn11r1}
%
\begin{bmisc}[auto:STB|2010-11-18|09:18:59]
\bauthor{\bsnm{McShane},~\bfnm{B.~B.}\binits{B.~B.}} \AND
\bauthor{\bsnm{Wyner},~\bfnm{A.~J.}\binits{A.~J.}}
(\byear{2011}b).
\bhowpublished{Supplement to ``Rejoinder on A statistical analysis of multiple temperature proxies:
Are reconstructions of surface temperatures over the last 1000 years reliable?''
DOI:
\href
{http://dx.doi.org/10.1214/10-AOAS398REJSUPPA}{10.1214/10-AOAS398REJSUPPA}}.
\end{bmisc}
%
\endbibitem

%b25 ###
\bibitem[\protect\citeauthoryear{McShane and Wyner}{2011c}]{McSWyn11r2}
%
\begin{bmisc}[auto:STB|2010-11-18|09:18:59]
\bauthor{\bsnm{McShane},~\bfnm{B.~B.}\binits{B.~B.}} \AND
\bauthor{\bsnm{Wyner},~\bfnm{A.~J.}\binits{A.~J.}}
(\byear{2011}c).
\bhowpublished{Supplement to ``Rejoinder on A statistical analysis of multiple temperature proxies:
Are reconstructions of surface temperatures over the last 1000 years reliable?''
DOI:
\href
{http://dx.doi.org/10.1214/10-AOAS398REJSUPPB}{10.1214/10-AOAS398REJSUPPB}}.
\end{bmisc}
%
\endbibitem

%b26 ###
\bibitem[\protect\citeauthoryear{NRC}{2006}]{NRC06}
%
\begin{bmisc}[auto:STB|2010-11-18|09:18:59]
\bauthor{NRC}
(\byear{2006}).
\bhowpublished{\textit{Surface Temperature Reconstructions}. The National
Academic Press, Washington, DC.
Available at \url{http://www.nap.edu/catalog.php?record\_id=11676}}.
\end{bmisc}
%
\endbibitem

%b27 ###
\bibitem[\protect\citeauthoryear{Rutherford et al.}{2005}]{Ruthetal05}
%
\begin{barticle}[auto:STB|2010-11-18|09:18:59]
\bauthor{\bsnm{Rutherford},~\bfnm{S.}\binits{S.}},
\bauthor{\bsnm{Mann},~\bfnm{M.~E.}\binits{M.~E.}},
\bauthor{\bsnm{Osborn},~\bfnm{T.~J.}\binits{T.~J.}},
\bauthor{\bsnm{Bradley},~\bfnm{R.~S.}\binits{R.~S.}},
\bauthor{\bsnm{Briffa},~\bfnm{K.~R.}\binits{K.~R.}},
\bauthor{\bsnm{Hughes},~\bfnm{M.~K.}\binits{M.~K.}} \AND
\bauthor{\bsnm{Jones},~\bfnm{P.~D.}\binits{P.~D.}}
(\byear{2005}).
\btitle{Proxy-based northern hemispheric surface reconstructions: Sensitivity
to method, predictor network, target season, and target doman}.
\bjournal{Journal of Climate}
\bvolume{18}
\bpages{2308--2329}.
\end{barticle}
%
\endbibitem

%b28 ###
\bibitem[\protect\citeauthoryear{Rutherford et al.}{2010}]{Ruthetal10}
%
\begin{barticle}[auto:STB|2010-11-18|09:18:59]
\bauthor{\bsnm{Rutherford},~\bfnm{S.}\binits{S.}},
\bauthor{\bsnm{Mann},~\bfnm{M.~E.}\binits{M.~E.}},
\bauthor{\bsnm{Ammann},~\bfnm{C.}\binits{C.}} \AND
\bauthor{\bsnm{Wahl},~\bfnm{E.}\binits{E.}}
(\byear{2010}).
\btitle{Comments on ``A surrogate ensemble study of climate reconstruction
methods: Stochasticity and robustness''.}
\bjournal{Journal of Climate}
\bvolume{23}
\bpages{2832--2838}.
\end{barticle}
%
\endbibitem

%b29 ###
\bibitem[\protect\citeauthoryear{Smerdon, Gonzalez-Rouco and
Zorita}{2008}]{SmeGonZor08}
%
\begin{bmisc}[auto:STB|2010-11-18|09:18:59]
\bauthor{\bsnm{Smerdon},~\bfnm{J.}\binits{J.}},
\bauthor{\bsnm{Gonzalez-Rouco},~\bfnm{J.~F.}\binits{J.~F.}} \AND
\bauthor{\bsnm{Zorita},~\bfnm{E.}\binits{E.}}
(\byear{2008}).
\bhowpublished{Comment on ``Robustness of proxy-based climate field
reconstruction methods'' by Michael E. Mann et al.
\textit{Journal of Geophysical Research --- Atmospheres} \textbf{113} D18106}.
\end{bmisc}
%
\endbibitem

%b30 ###
\bibitem[\protect\citeauthoryear{Smerdon and Kaplan}{2007}]{SmeKap07}
%
\begin{bmisc}[auto:STB|2010-11-18|09:18:59]
\bauthor{\bsnm{Smerdon},~\bfnm{J.}\binits{J.}} \AND
\bauthor{\bsnm{Kaplan},~\bfnm{A.}\binits{A.}}
(\byear{2007}).
\bhowpublished{Comment on ``Testing the Fidelity of methods used in proxy-based
reconstructions of past climate'': The role of the standardization
interval,
by M. E. Mann, S. Rutherford, E. Wahl, and C. Ammann.
\textit{Journal of Climate} \textbf{20} 5666--5670}.
\end{bmisc}
%
\endbibitem

%b31 ###
\bibitem[\protect\citeauthoryear{Smerdon, Kaplan and Chang}{2008}]{SmeKapCha08}
%
\begin{barticle}[auto:STB|2010-11-18|09:18:59]
\bauthor{\bsnm{Smerdon},~\bfnm{J.}\binits{J.}},
\bauthor{\bsnm{Kaplan},~\bfnm{A.}\binits{A.}} \AND
\bauthor{\bsnm{Chang},~\bfnm{D.}\binits{D.}}
(\byear{2008}).
\btitle{On the origin of the standardization sensitivity in regem
climate field
reconstructions}.
\bjournal{Journal of Climate}
\bvolume{21}
\bpages{6710--6723}.
\end{barticle}
%
\endbibitem

%b32 ###
\bibitem[\protect\citeauthoryear{Smerdon, Kaplan and
Amrhein}{2010}]{SmeKapAmr10}
%
\begin{barticle}[auto:STB|2010-11-18|09:18:59]
\bauthor{\bsnm{Smerdon},~\bfnm{J.}\binits{J.}},
\bauthor{\bsnm{Kaplan},~\bfnm{A.}\binits{A.}} \AND
\bauthor{\bsnm{Amrhein},~\bfnm{D.~E.}\binits{D.~E.}}
(\byear{2010}).
\btitle{Erroneous model field representations in multiple pseudoproxy studies:
Corrections and implications}.
\bjournal{Journal of Climate}
\bvolume{23}
\bpages{5548--5554}.
\end{barticle}
%
\endbibitem

%b33 ###
\bibitem[\protect\citeauthoryear{Smerdon et al.}{2010}]{SmKaChEv10}
%
\begin{barticle}[auto:STB|2010-11-18|09:18:59]
\bauthor{\bsnm{Smerdon},~\bfnm{J.}\binits{J.}},
\bauthor{\bsnm{Kaplan},~\bfnm{A.}\binits{A.}},
\bauthor{\bsnm{Chang},~\bfnm{D.}\binits{D.}} \AND
\bauthor{\bsnm{Evans},~\bfnm{M.}\binits{M.}}
(\byear{2010}).
\btitle{A pseudoproxy evaluation of the cca and regem methods for
reconstructing climate fields of the last millennium}.
\bjournal{Journal of Climate}
\bvolume{23}
\bpages{4856--4880}.
\end{barticle}
%
\endbibitem

%b34 ###
\bibitem[\protect\citeauthoryear{Sundberg}{1999}]{Sun99}
%
\begin{barticle}[mr]
\bauthor{\bsnm{Sundberg},~\bfnm{Rolf}\binits{R.}}
(\byear{1999}).
\btitle{Multivariate calibration---direct and indirect regression methodology}.
\bjournal{Scand. J. Statist.}
\bvolume{26}
\bpages{161--207}.
\bid{doi={10.1111/1467-9469.00144}, mr={1707599}}
\end{barticle}
%
\endbibitem

%b35 ###
\bibitem[\protect\citeauthoryear{ter Braak}{1995}]{terBr95}
%
\begin{barticle}[auto:STB|2010-11-18|09:18:59]
\bauthor{\bparticle{ter} \bsnm{Braak},~\bfnm{C.}\binits{C.}}
(\byear{1995}).
\btitle{Non-linear methods for multivariate statistical calibration and their
use in palaeoecology: A comparison of inverse ($k$-nearest neighbours,
partial least squares and weighted averaging partial least squares) and
classical approaches}.
\bjournal{Chemometrics and Intelligent Laboratory Systems}
\bvolume{28}
\bpages{165--180}.
\end{barticle}
%
\endbibitem

%b36 ###
\bibitem[\protect\citeauthoryear{Tingley and Huybers}{2010}]{TinHuy10a}
%
\begin{barticle}[auto:STB|2010-11-18|09:18:59]
\bauthor{\bsnm{Tingley},~\bfnm{M.}\binits{M.}} \AND
\bauthor{\bsnm{Huybers},~\bfnm{P.}\binits{P.}}
(\byear{2010}).
\btitle{A Bayesian algorithm for reconstructing climate anomalies in
space and
time. Part I: Development and applications to paleoclimate reconstruction
problems}.
\bjournal{Journal of Climate}
\bvolume{23}
\bpages{2759--2781}.
\end{barticle}
%
\endbibitem
\

%b37 ###
\bibitem[\protect\citeauthoryear{Zou and Hastie}{2005}]{ZouHas05}
%
\begin{barticle}[mr]
\bauthor{\bsnm{Zou},~\bfnm{Hui}\binits{H.}} \AND
\bauthor{\bsnm{Hastie},~\bfnm{Trevor}\binits{T.}}
(\byear{2005}).
\btitle{Regularization and variable selection via the elastic net}.
\bjournal{J. R. Stat. Soc. Ser. B Stat. Methodol.}
\bvolume{67}
\bpages{301--320}.
\bid{doi={10.1111/j.1467-9868.2005.00503.x}, mr={2137327}}
\end{barticle}
%
\endbibitem

\end{thebibliography}
\end{document}